\documentclass[12pt,preprint]{aastex}
\usepackage{natbib}
\usepackage{graphicx}
\begin{document}
\title{Luminous Blue Variable Stars In The Two Extremely Metal-Deficient Blue Compact Dwarf
Galaxies DDO 68 and PHL 293B
}
\author{Yuri I. Izotov}
\affil{Main Astronomical Observatory, Ukrainian National Academy of Sciences,
27 Zabolotnoho str., Kyiv 03680, Ukraine}
\email{izotov@mao.kiev.ua}
\and
\author{Trinh X. Thuan}
\affil{Astronomy Department, University of Virginia, P.O. Box 400325, 
Charlottesville, VA 22904-4325}
\email{txt@virginia.edu}

\begin{abstract}
We present photometric and spectroscopic observations 
of two luminous blue variable (LBV) stars in two extremely metal-deficient
blue compact dwarf (BCD) galaxies, DDO 68 with 12+logO/H = 7.15 and PHL 293B with  
12+logO/H = 7.72. These two BCDs are the lowest-metallicity galaxies where
LBV stars have been detected, allowing to study the LBV 
phenomenon in the extremely low metallicity regime, and shedding light 
of the evolution of the first generation of massive stars born from 
primordial gas. 
We find that the strong outburst of the LBV star in
DDO 68 occurred sometime between February 2007 and January 2008.
We have compared the properties of the broad line emission in low-metallicity LBVs
with those in higher metallicity LBVs. We find that, for the LBV
star in DDO 68, broad emission with 
a P Cygni profile is seen in both H and He {\sc i} emission lines.
On the other hand, for the LBV star
in PHL 293B, P Cygni profiles are detected only in H lines. 
For both LBVs, no heavy element emission line such as Fe {\sc ii} 
was detected.
The H$\alpha$ luminosities of LBV stars in both galaxies are comparable 
to the one obtained for the 
LBV star in NGC 2363 (Mrk 71) which has a higher metallicity 12+logO/H = 7.89.
On the other hand, the terminal velocities of the stellar winds in 
both low-metallicity 
LBVs are high, 
$\sim$ 800 km s$^{-1}$, a factor of $\sim$ 4 
higher than the terminal velocities of the winds in high-metallicity LBVs. 
This suggests that stellar winds at low metallicity are driven by 
a different mechanism than the one operating in high-metallicity winds.
\end{abstract}


\section{INTRODUCTION}

The most massive stars evolve on very short timescales of a few million years
compared to the very long time scales of billion years of solar-type stars. 
Thus, important evolutionary events in the life of very massive stars such as 
core-collapse supernovae or giant eruptions of luminous blue variables 
\citep[LBVs; ][]{C84} are hard to catch and study. To maximize the probability 
of catching such an event, it is best to observe objects known to contain 
many massive stars. Blue compact dwarf (BCD) galaxies are ideal  
laboratories for such studies as they are undergoing 
intense bursts of star formation and their star-forming regions
are known to harbor up to tens of thousands 
of massive stars \citep{TI05}. Furthermore, mass loss is a critical 
factor in determining the evolution of a 
massive star. The mass loss rate depends in turn on the metallicity of the gas, 
if driven by line opacity. BCDs are also excellent laboratories to 
test metallicity effects on mass loss as they are metal-deficient, their 
metallicity ranging from 
about half to a few percent the Sun's metallicity \citep{I07}. 
Thus BCDs are the only 
galaxies in the local Universe where low-metallicity massive stars do exist. 
They are unique objects for studying processes related to massive star 
formation and evolution and their interaction
with the interstellar medium, and for testing models
of massive stellar evolution. They are the best local approximations to 
primordial galaxies.  

BCDs have been used to investigate mass loss through 
observational signatures of     
stellar winds from massive stars. The prevailing belief is 
 that the efficiencies of stellar winds in massive stars
are significantly reduced at low metallicities. However, this conclusion is based
mainly on theoretical extrapolations and observations of relatively high-metallicity 
massive stars in our Galaxy (12 + log O/H $\sim$ 8.7), the Large Magellanic 
Cloud (12 + log O/H $\sim$ 8.4)
and the Small Magellanic Cloud (12 + log O/H $\sim$ 8.1) and some
other nearby high-metallicity galaxies. There is growing
evidence suggesting that massive stars do have stellar winds, 
even in the most metal-deficient BCDs known, with oxygen abundances 
12 + log O/H $\la$ 7.6. Thus, \citet{I97} and \citet{L97} discovered
a Wolf-Rayet (WR) stellar population in I Zw 18, the most metal-deficient 
emission-line galaxy known at the time, with oxygen abundance 
12+logO/H = 7.17 $\pm$ 0.01 \citep{I99}. \citet{GIT00} have shown that the
observed ratio of the number of WR stars to the number of O stars
in I Zw18 cannot be reproduced by theoretical population synthesis models 
based
on non-rotating stellar evolution models and observational properties of 
high-metallicity WR stars. To achieve agreement between observations and theory, 
  \citet{CH06} have suggested that 
WR stars in I Zw 18 have lower emission-line luminosities than 
their high-metallicity counterparts.
Furthermore, \citet{TI97} have observed P Cygni profiles for the Si {\sc iv} $\lambda$1394, 
$\lambda$1403 absorption lines in the spectrum of the
emission-line galaxy SBS 0335--052E, one of the most metal-deficient BCD known,
 with 12 + log O/H =
7.30 $\pm$ 0.01 \citep{I07}. It thus appears that extremely 
low-metallicity massive stars do possess stellar winds, although
the properties of these stellar winds may significantly differ
from those of their high-metallicity counterparts.

To find more objects with observational signatures of massive star 
activity, we have recently assembled 
a large sample of about 40 emission-line dwarf galaxies which exhibit 
broad components 
in their strong emission lines, mainly in H$\beta$, 
[O {\sc iii}] $\lambda\lambda$ 4959, 5007, and 
H$\alpha$ \citep{I07}. Except for four objects which appear to 
contain intermediate-mass black holes \citep{IT08}, 
the broad emission of all the other objects can be 
attributed to some evolutionary stage of massive stars and to 
their interaction with the circumstellar and interstellar medium:
WR stars, supernovae, superbubbles or LBV stars.
In particular, our attention was drawn to the peculiar spectrum of the BCD 
PHL 293B $\equiv$ SDSS J2230--0006, with an oxygen abundance 
12 + log O/H = 7.66 $\pm$ 0.04, which shows broad hydrogen emission lines
with P Cygni profiles, spectral features that are characteristic of a  
LBV star. Coincidentally, several months after our finding, \citet{P08} discovered 
another bright LBV in the BCD DDO 68. This BCD,
with an oxygen abundance 12 + log O/H = 7.14 $\pm$ 0.03 is even 
more metal-deficient than PHL 293B \citep{IT07}. It is the second most metal-deficient 
star-forming galaxy known, after SBS 0335--052W.   

The study of these newly discovered LBV stars in the two 
very metal-deficient galaxies DDO 68 and PHL 293B is 
the focus of this paper.
Giant eruptions of LBV stars, with a brightening greater than 3 mag, 
are exceedingly rare. In the Galaxy, although there
are some 35 confirmed or candidate LBVs \citep{C05}, only two,  
$\eta$ Carinae  and P Cygni, are known to have undergone
such giant eruptions over the last 5 centuries, $\eta$ Carinae between 1837 
and 1860, and P Cygni, twice in the 17th century. P Cygni was discovered 
in 1600 as a naked eye star, and remained bright for many years, before 
fading and re-appearing in 1655, and then fading again.
Typically, LBVs exhibit day-to-day microvariations in brightness
of 0.1-0.2 mag, and ``normal'' irregular variations of 
1--2 mag on a timescale of years, in which the spectral type may vary from 
an early B supergiant
at visual minimum to a late B or early A supergiant at visual maximum
\citep[e.g.][]{C84,HD94,L94,N96,D97,D01,C04,P06,W08}.
There have been also LBVs observed in the Large Magellanic Cloud 
and other Local Group galaxies \citep{M07}. 
 
The LBV phase is believed to represent a critical transition in the late 
evolution of all stars with initial masses greater than about 50 $M_\odot$. 
During this phase, the stars lose sufficiently large amounts of mass in recurrent 
explosive events, to go from the stage of O stars burning hydrogen on the 
main-sequence to that of core-helium burning classical WR stars. In the 
HR diagram, LBVs lie just to the left of the Humphreys-Davidson limit
\citep{HD94},
beyond which no stars are observed. They appear to define the locus of an 
instability that prevents further redward evolution.  

The understanding of the physical processes operating during the 
LBV phase is thus crucial for modeling massive star evolution. However, 
the exact mechanism giving rise to a LBV outburst is still unknown, even if 
many LBVs at solar metallicity have been studied. The situation is worse
at low metallicity because
the only relatively low-metallicity LBV star that has been 
observationally investigated in detail 
is the star V1 in the H {\sc ii} region NGC 2363 (Mrk 71) of  
the dwarf cometary galaxy NGC 2366 \citep{D97,D01,P06}, with 
12+log O/H = 7.89 \citep{G94,ITL97}. High-quality spectroscopic 
studies of LBV stars
 in two more extremely metal-deficient galaxies should help us to  
better understand how the LBV phase depends on metallicity.    

We present here new spectroscopic observations of the
LBVs in DDO 68 and PHL 293B and derive their emission line 
parameters. We also discuss new photometric observations 
of DDO 68 to better constrain the epoch of the LBV outburst in it.
The observations are described in \S2. 
We discuss in \S3 the photometric variations of the 
two LBV stars. We derive the element abundances in 
the two host galaxies and show how  
the main properties of the stellar winds in low-metallicity LBVs
differ significantly from those in their high-metallicity counterparts. 
Our conclusions are summarized in \S4.

\section{OBSERVATIONS}

\subsection{Imaging}

New deep images of DDO 68 were obtained with the 
SDSS $g$ and $i$ filters \citep{F96} using the 2.1m 
telescope at Kitt Peak National Observatory\footnote{Kitt Peak National 
Observatory is operated by 
the Association of Universities for Research in Astronomy (AURA) 
under cooperative agreement with the National Science Foundation.}
on the night 9 February 2007. The details of the observations are shown in 
Table \ref{tab1}. These observations were compared to the SDSS images of DDO 68 
in the same filters obtained on 16 April 2004. The details of these observations 
are also shown in Table \ref{tab1}. To process the 2.1m KPNO images we have 
used the standard
data reduction procedures in IRAF\footnote{IRAF is distributed by National 
Optical Astronomy 
Observatory, which is operated by the Association of Universities for 
Research in Astronomy, Inc., under cooperative agreement with the National 
Science Foundation.}, which include bias subtraction and flat-field 
correction. For photometric calibration, we have used bright stars in the field of 
SDSS images. The resulting $g$ and $g-i$ 2.1m KPNO 
images of DDO 68 are shown in Fig. \ref{fig1}, with the 
most prominent H {\sc ii} regions labeled, following the 
numbering system of \citet{P05}. It is seen
from the $g-i$ image that the light of DDO 68 is dominated by blue young stellar
populations (in black) without evidence for a red halo. This suggests that
DDO 68 may be a young system, with an age less than 1 Gyr, 
  as discussed by \citet{P07}. The several compact
red objects in the image (in white) are either 
foreground Galactic stars or distant background galaxies.

\subsection{Spectroscopy}

\begin{figure*}
\figurenum{1}
\hbox{\includegraphics[angle=0,width=0.5\linewidth]{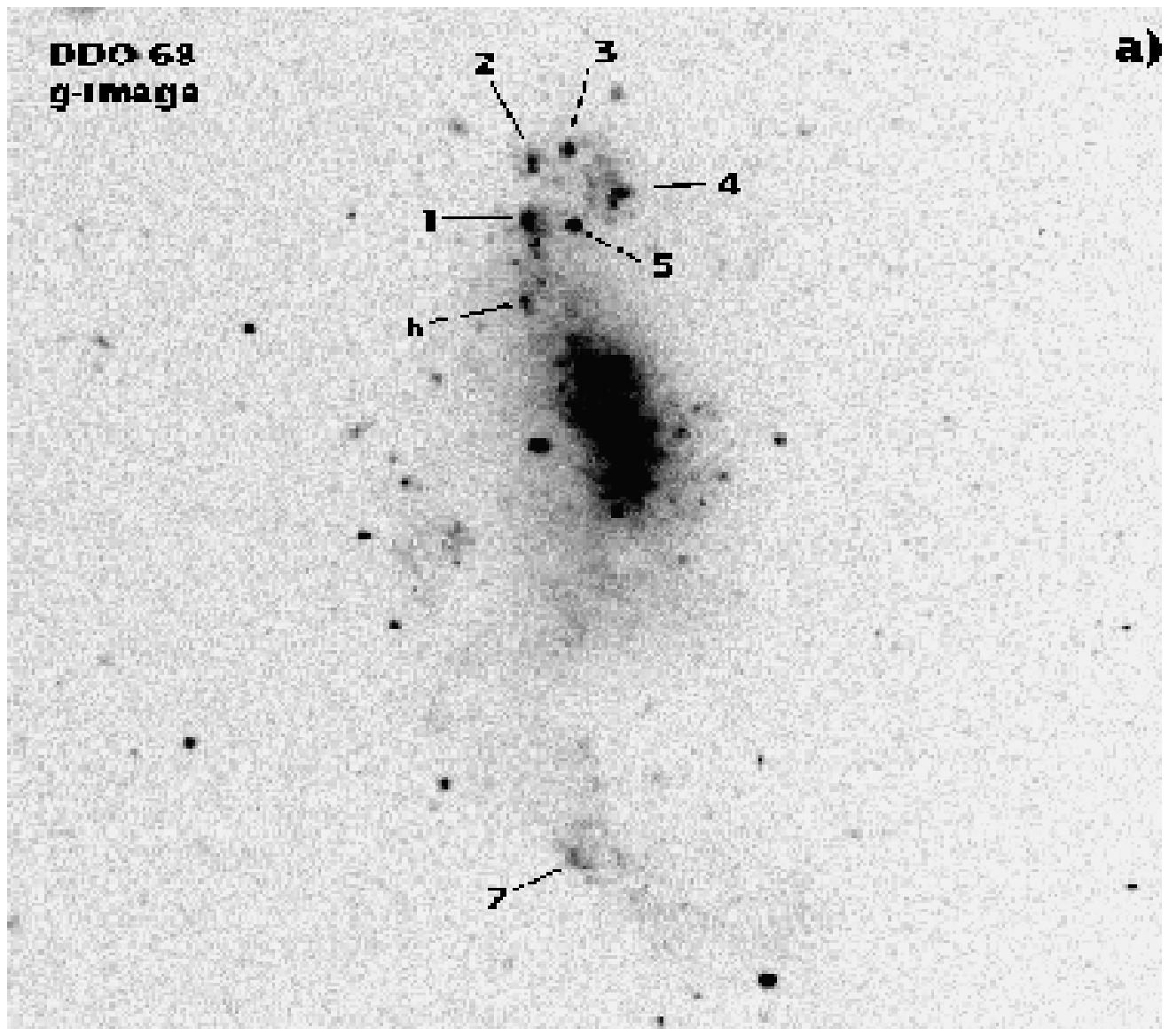} 
\includegraphics[angle=0,width=0.5\linewidth]{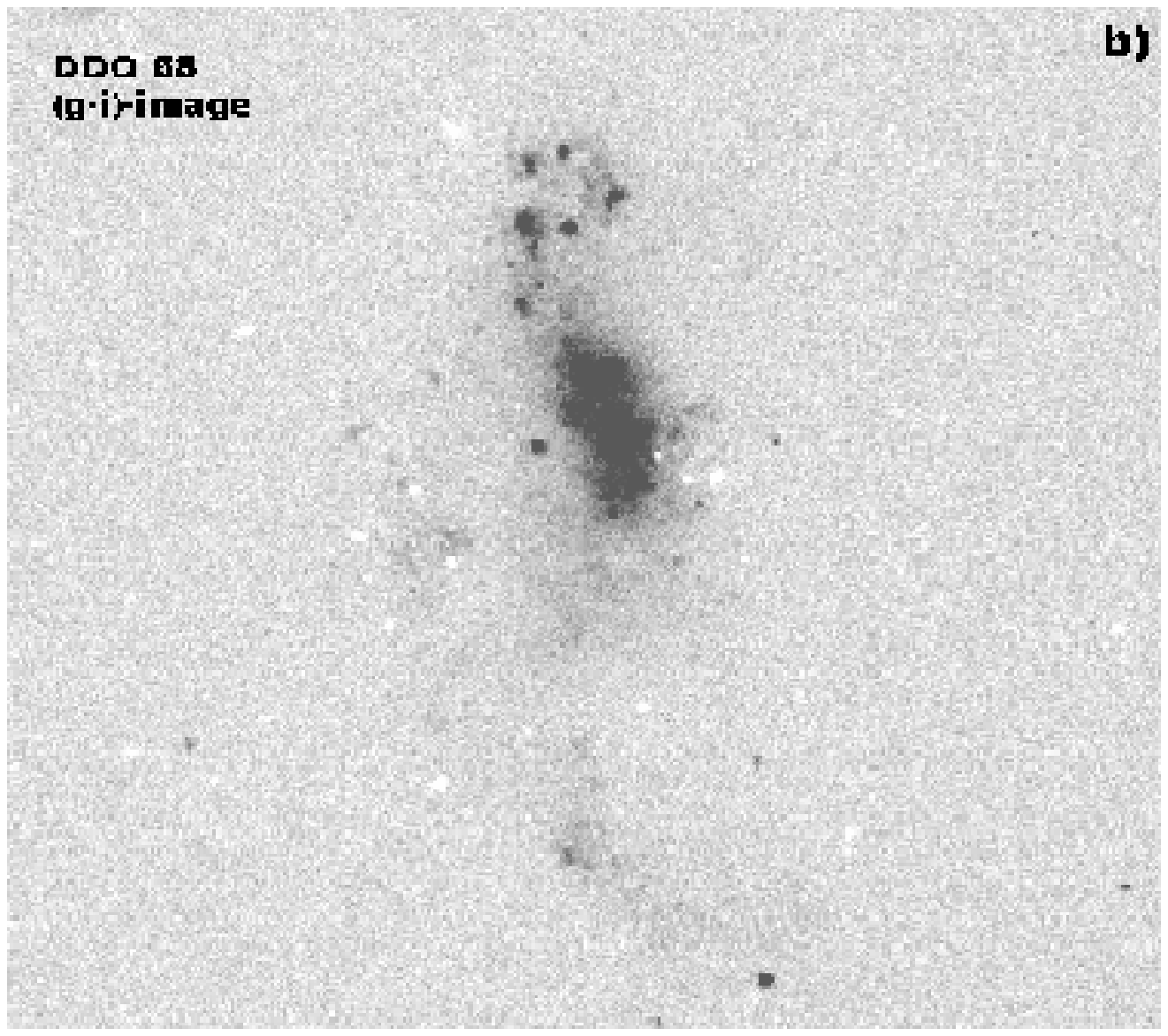}
}
\figcaption{a) 2\farcm5$\times$2\farcm5 2.1m KPNO $g$ image of DDO 68. The H {\sc ii} 
regions are labeled
following \citet{P05}. b) 2.1m KPNO $g-i$ image of DDO 68.
\label{fig1}}
\end{figure*}

\begin{figure*}
\figurenum{2}
\hbox{\includegraphics[angle=0,width=0.5\linewidth]{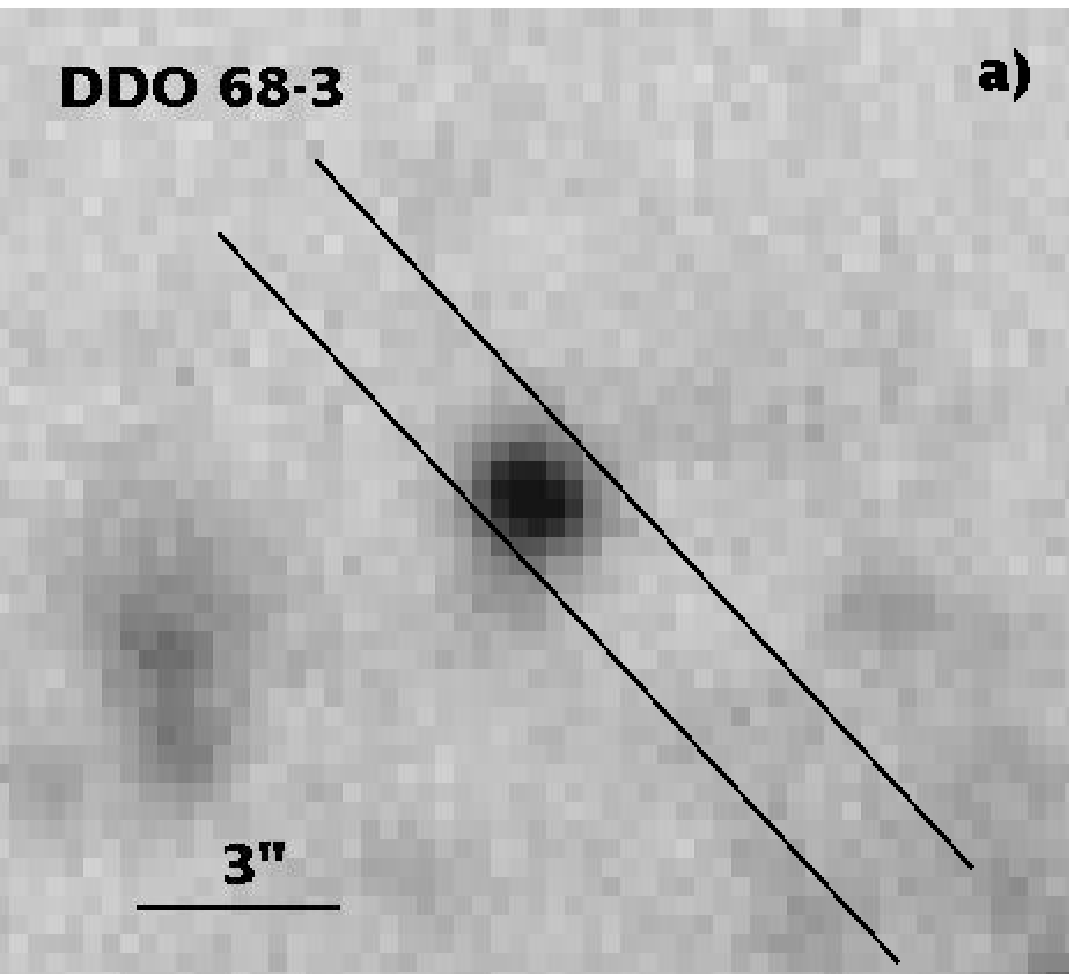}
\includegraphics[angle=0,width=0.5\linewidth]{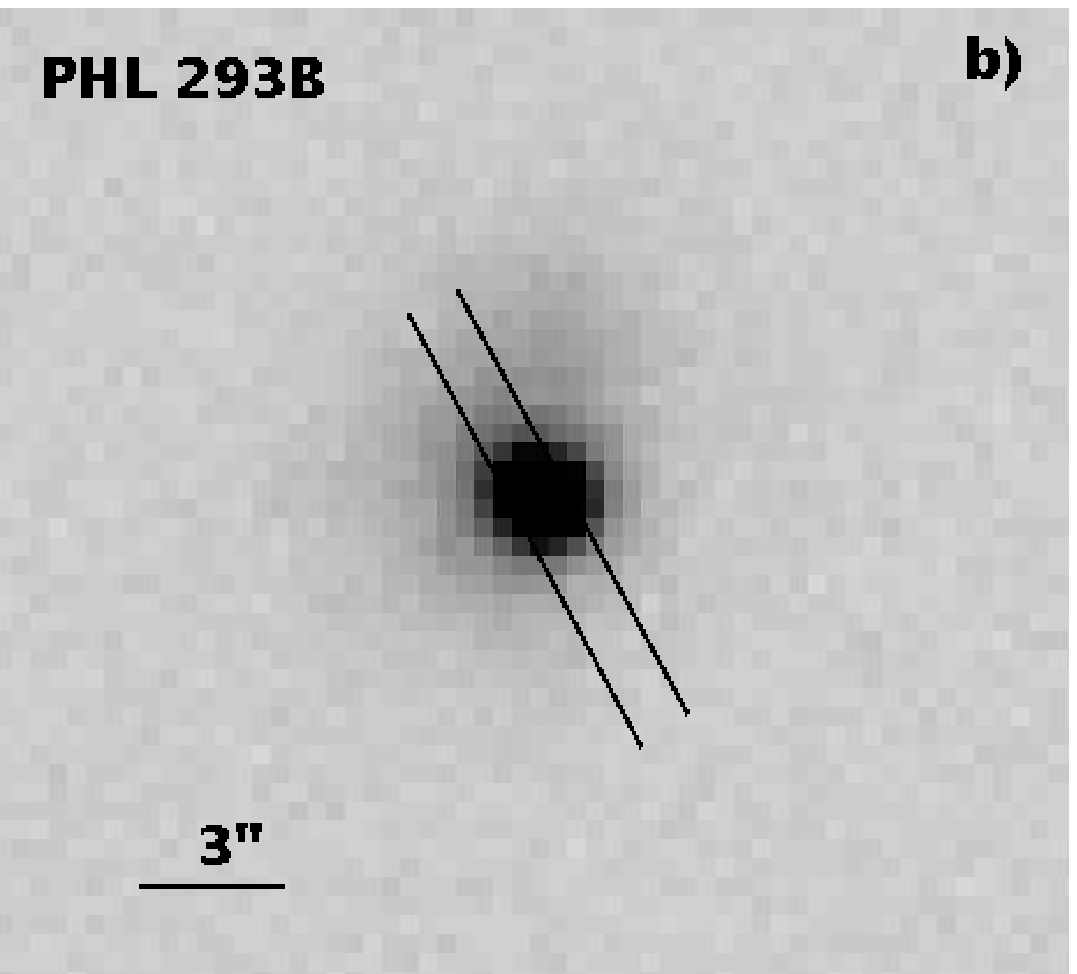}
}
\hbox{\includegraphics[angle=-90,width=0.5\linewidth]{f2c.ps}
\includegraphics[angle=-90,width=0.5\linewidth]{f2d.ps}
}
\figcaption{a) MMT slit superposed on a KPNO 2.1m $g$ band image of DDO 68-3. 
b) VLT slit superposed on a SDSS $g$ band image of PHL 293B.
 c) Normalized H$\beta$ brightness distribution (solid line) and 
normalized continuum distribution (dashed line)
of DDO 68-3 along the MMT slit. d) The same brightness distributions as in c)
but in PHL 293B along the VLT slit. \label{fig2}}
\end{figure*}

New high signal-to-noise ratio 
optical spectra of DDO 68 were obtained
using the 3.5 m Apache Point Observatory (APO)\footnote{The Apache Point Observatory 
3.5-meter telescope is owned and operated by the Astrophysical Research 
Consortium.} telescope and the 6.5 m MMT\footnote{The MMT is operated by 
the MMT Observatory (MMTO), a joint venture of the Smithsonian Institution  
and the University of Arizona.} on the nights of 2 February 2008 and
28 March 2008, respectively. 
The APO observations of DDO 68 were made with the Dual Imaging Spectrograph (DIS),
which covers both the blue and red wavelength ranges. A 1\farcs5$\times$360\arcsec\ 
slit was used. In the blue range, we use the B400 grating with a linear dispersion
of 1.83 \AA/pix and a central wavelength of 4400\AA,  while in the red range
we use the R300 grating with a linear dispersion of 2.31 \AA/pix and 
a central wavelength of 7500\AA. The above instrumental set-up gave a spatial 
scale along the slit of 0\farcs4 pixel$^{-1}$, a spectral range of 
$\sim$3600 -- 9600\AA\ and a spectral resolution of 7\AA\ (FWHM).
The slit was oriented in such a way as to include both 
regions 3 and 4 (hereafter DDO 68-3 and DDO 68-4) (Figure \ref{fig1}).
The MMT observations of DDO 68 were made with the Blue Channel spectrograph. We used
a 1\farcs5$\times$180\arcsec\ slit and a 800 grooves/mm grating in first order.
The above instrumental set-up gave a spatial scale along the slit of 0\farcs6
pixel$^{-1}$, a scale perpendicular to the slit of 0.75\AA\ pixel$^{-1}$,
a spectral range 3200 -- 5000\AA\ and a spectral resolution of 3\AA\ (FWHM).
We show the MMT slit location superposed on an expanded image of DDO 68-3 
in Fig. \ref{fig2}a. The star-forming region is nearly unresolved. 
However, the brightness distribution along the slit suggests that 
the LBV star is contained within it (Fig. \ref{fig2}c). 

As for PHL 293B, we did not obtain new observations but 
used a high spectral resolution spectrum of PHL 293B in the 
archives of the European Southern Observatory (ESO). 
The spectrum was obtained with the UVES spectrograph on the 8m 
VLT\footnote{Based on 
observations made with ESO Telescopes at the La Silla Paranal Observatory 
under programme ID 70.B-0717(A)}  during the night of
8 November 2002, in both the blue arm (grating CD\#2, central
wavelength 3900\AA, slit 1\arcsec$\times$8\arcsec) and the  
red arm modes (grating CD\#3, central
wavelength 5800\AA, slit 1\arcsec$\times$12\arcsec), giving a  
wavelength range 3100 -- 6800\AA. The resolving power was
$\sim$ 80,000, resulting in a spectral resolution $\sim$ 0.2\AA. 
The VLT slit superposed on a $g$ band SDSS image of PHL 293B is shown in 
Fig. \ref{fig2}b. As in the case of DDO 68-3,  
the brightness distribution along the slit (Fig. \ref{fig2}d)
suggests that the LBV star resides within the bright, compact and barely 
resolved star-forming region. 
The ESO 
spectrum of PHL 293B was supplemented by the SDSS spectrum, obtained on 
22 August 2001, with a wavelength range 3800 -- 9200\AA\ and a spectral 
resolution of $\sim$ 3\AA. The details of the spectroscopic observations 
are also 
given in Table \ref{tab1}.

For both APO and MMT observations, several Kitt Peak and CTIO IRS spectroscopic 
standard stars were observed for flux
calibration. Spectra of He-Ne-Ar comparison arcs were obtained 
at the beginning of each night for wavelength calibration. 
As for the high spectral resolution VLT/UVES observations, 
calibration was done by observing the standard star Feige 110, and wavelength 
calibration was done using the spectrum of a Thorium comparison arc. 

The data reduction procedures of the spectroscopic observations
are the same as described in \citet{IT07}.
Briefly, the two-dimensional spectra were first bias subtracted and 
flat-field corrected with IRAF.
We then use the IRAF
software routines IDENTIFY, REIDENTIFY, FITCOORD, TRANSFORM to 
perform wavelength
calibration and correct for distortion and tilt for each frame. 
Background subtraction was performed using the routine BACKGROUND. The level of
background emission was determined from the closest regions to the galaxy 
that are free of galaxian stellar and nebular line emission,
 as well as of emission from foreground and background sources.
A one-dimensional spectrum was then extracted from the two-dimensional 
frame using the APALL routine. 
Before extraction, separate two-dimensional spectra of each object 
were carefully aligned using the location of the brightest part in
each spectrum, so that spectra were extracted at the same positions in all
subexposures. We then summed the individual spectra 
from each subexposure after manual removal of the cosmic rays hits. 

\begin{figure*}
\figurenum{3}
\hbox{\includegraphics[angle=-90,width=1.0\linewidth]{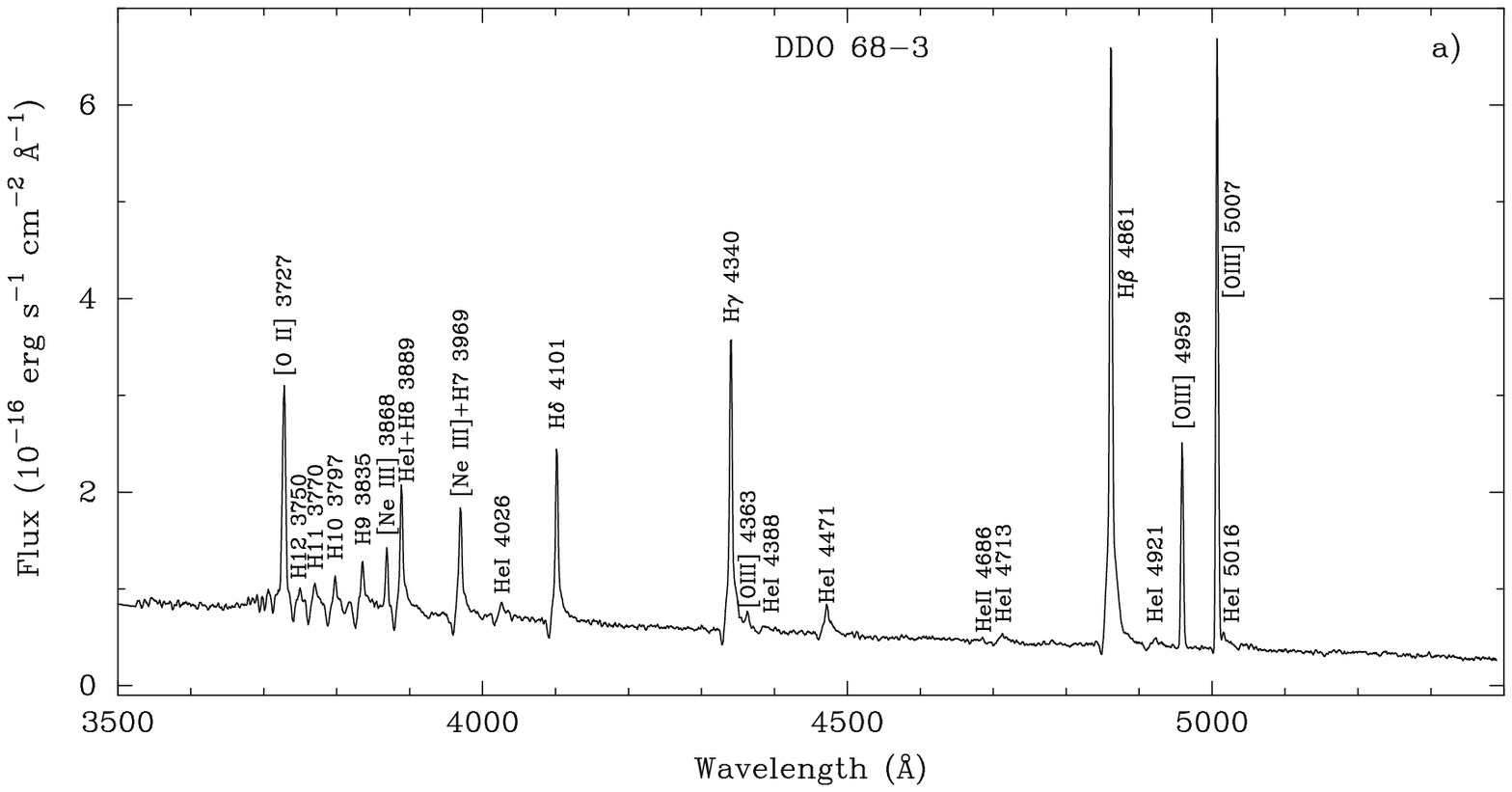}} 
\hbox{\includegraphics[angle=-90,width=1.0\linewidth]{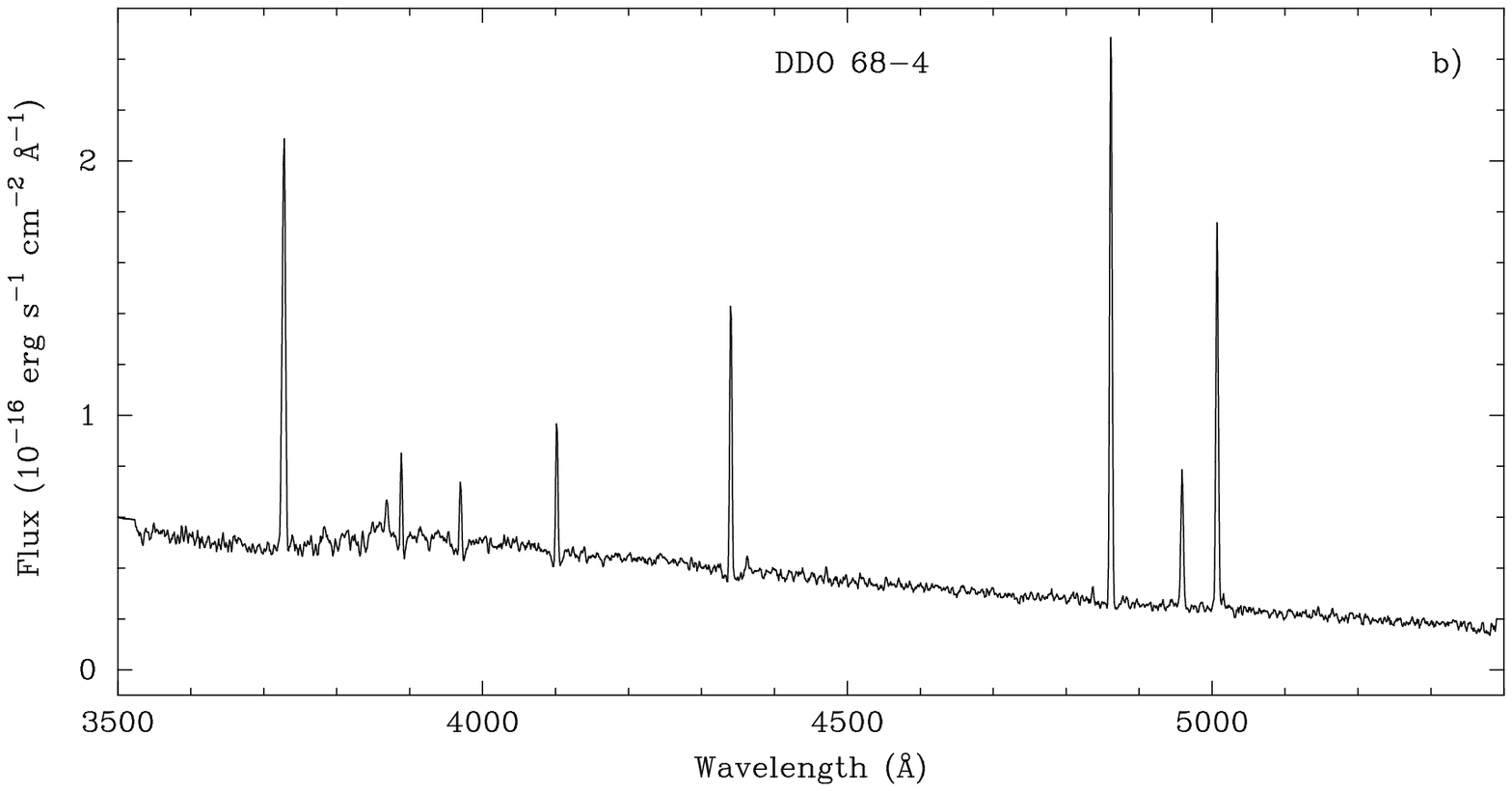}
}
\figcaption{Redshift-corrected 6.5 MMT 
spectra of DDO 68-3 (a) DDO 68-4 (b). The emission lines are labeled in 
panel a). \label{fig3}}
\end{figure*}

The resulting MMT spectra of DDO 68-3 and DDO 68-4 are shown in 
Fig. \ref{fig3}. The spectrum of DDO 68-3 is strikingly different 
from normal spectra of H {\sc ii} regions: it shows  
strong broad hydrogen and helium emission lines with P Cygni profiles
superposed on narrow nebular
emission lines (Fig. \ref{fig3}a). No emission lines of heavy elements 
(e.g., Fe {\sc ii} and N {\sc iii} lines) were detected, in contrast to 
high-metallicity quiescent LBV spectra which show the presence of many such lines. 
The absence of permitted heavy 
element emission lines in the visible spectrum of the LBV star 
in DDO 68-3 is most 
likely due to its very low metallicity, but probably  
also to temperature effects 
in the pseudo-photosphere of the LBV star as it is likely in an eruptive 
phase.  
No broad emission is detected in DDO 68-4 (Fig. \ref{fig1}b). 
The [O {\sc iii}] $\lambda$4363 emission line is detected in both DDO 68-3 
and DDO 68-4, 
allowing a direct determination of the electron temperature and 
element abundances.

\begin{figure*}
\figurenum{4}
\hbox{\includegraphics[angle=-90,width=1.0\linewidth]{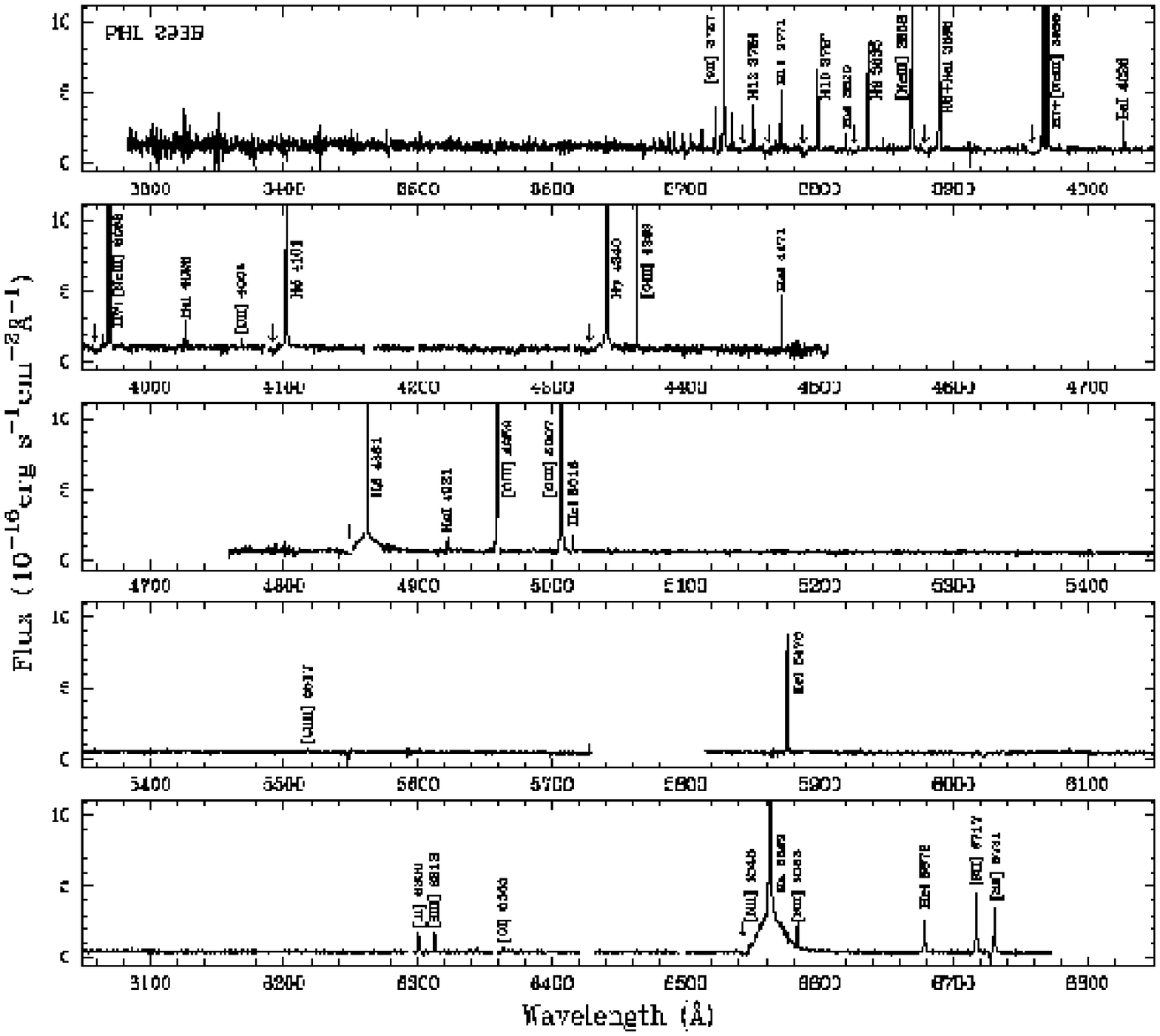} 
}
\figcaption{Redshift-corrected archival 8m VLT/UVES spectrum of PHL 293B. The locations of blue-shifted 
absorption in lines with P Cygni profiles are shown by arrows.
\label{fig4}}
\end{figure*}

The high-resolution 
VLT spectrum of PHL 293B is shown in Fig. \ref{fig4}. Broad emission and 
P Cygni profiles are clearly seen for the H$\alpha$,
H$\beta$ and H$\gamma$ lines,  
superposed on narrow nebular emission lines.
The broad emission is not evident in the higher-order hydrogen lines, even if 
blue absorption is present (marked by arrows in Fig. \ref{fig4}). However, at variance with
DDO 68, no broad He {\sc i} emission is seen in the spectrum of
PHL 293B. Similar to DDO 68, no permitted heavy element emission lines 
with P Cygni profiles were detected in the spectrum of PHL 293B, 
again a consequence of the low metallicity of the LBV star.

\section{RESULTS AND DISCUSSION}

\subsection{Photometry of the LBV star in DDO 68}

We now examine the temporal variation of the brightness of the 
LBV star in DDO 68, in 
an attempt to narrow down the time of its outburst. 
It was first seen by \citet{P08} thanks to the presence 
of broad emission in the spectrum
of DDO 68-3 obtained on 11 January 2008. Earlier spectroscopic 
observations 
of the same region by \citet{P05} in 2005 did not reveal anything 
abnormal. This implies that outburst occurred sometime during the 
2005 -- 2007 period.
We can use the SDSS and our 2.1m KPNO imaging observations of DDO 68 
to narrow down this range further. With the IRAF APPHOT package, we have 
measured on both sets of images
$g$ and $i$ magnitudes of the H {\sc ii} regions in DDO 68 
that are labeled in Fig. \ref{fig1}a. The SDSS images were obtained 
on 16 April 2004 and the KPNO images on 
9 February 2007. The results of our photometric measurements are shown in 
Table \ref{tab2}. Examination of this table shows that the 
$g$ magnitudes of regions 1 -- 6 derived from KPNO 
and SDSS images are in very good agreement when measurement errors are 
taken into account. 
  Thus, 
the $g$ magnitude difference between the two sets of images 
for DDO 68-3 (where the LBV star has been identified spectroscopically) 
is only -0.06 mag and can be attributed 
to photometric errors. The agreement between the $i$ magnitudes is not
as good, primarily because of the noisy short exposure SDSS $i$ image.
We thus conclude that no drastic luminosity change has occurred in 
DDO 68-3 up to 9 February 2007, and that the LBV 
outburst has occurred sometime between 9 February 2007 and 11 January 2008.

As for PHL 293B, except for the SDSS images, 
we do not possess other images of it taken at subsequent times.
All we can say is that the LBV outburst in it occurred prior to 
22 August 2001, the date of the SDSS spectrum, as it already showed broad 
emission by then.   
Clearly, imaging of both DDO 68 and PHL 293B is urgently needed to constrain the 
magnitude jump of the LBV outburst in the two BCDs. Are both eruptions 
``normal'' LBV outbursts with $\delta$$m$ $\sim$ 1-2 mag, or major outbursts 
with  $\delta$$m$ $\ge$ 3 mag?  

\subsection{Element abundances}

We now derive element abundances for both objects 
from the narrow emission line fluxes.
These fluxes have been 
measured using Gaussian fitting with the IRAF SPLOT routine. 
They have been corrected for both extinction, using the reddening curve
of \citet{W58}, and underlying
hydrogen stellar absorption, derived simultaneously by an iterative procedure as
described by \citet{ITL94} and using the observed decrements of the 
narrow hydrogen Balmer lines. It is assumed in
this procedure that hydrogen line emission is produced only by 
spontaneous transitions
in recombination cascades, i.e. we neglect possible collisional excitation.
Such a situation usually holds in low-density H {\sc ii} regions ionized by 
stellar radiation such as those considered here. 
The extinction-corrected fluxes 
100$\times$$I$($\lambda$)/$I$(H$\beta$) of 
the narrow lines for each galaxy, together with the extinction coefficient
$C$(H$\beta$), the equivalent width of the H$\beta$ emission line 
EW(H$\beta$), the H$\beta$ observed flux $F$(H$\beta$) and the 
equivalent widths of the 
underlying hydrogen absorption lines EW(abs) are given in Table \ref{tab3}.
The physical conditions and element abundances of DDO 68-3 (APO and MMT
observations) and DDO 68-4 (MMT observations) and of the H {\sc ii}
region in PHL 293B (VLT observations) are derived from the narrow line fluxes 
following \citet{I06a}. Briefly, we adopt a two-zone photoionized H {\sc ii} 
region model: a high-ionization zone with electron 
temperature $T_e$([O {\sc iii}]),
where O {\sc iii}, Ne {\sc iii}, and Ar {\sc iv} originate, and 
a low-ionization zone with electron temperature $T_e$([O {\sc ii}]), where 
O {\sc ii}, N {\sc ii}, S {\sc ii} and Fe {\sc iii} originate. 
The temperature $T_e$([O {\sc iii}])
is derived from the [O {\sc iii}] $\lambda$4363/($\lambda$4959+$\lambda$5007)
flux ratio. This temperature is used for the determination of the O$^{2+}$
and Ne$^{2+}$ ionic abundances. The temperature $T_e$([O {\sc ii}]) is  
obtained from the relation between $T_e$([O {\sc ii}] and $T_e$([O {\sc iii}]
derived by \citet{I06a} by fitting photoionization models.  
This temperature is used for the determination of the 
O$^+$ and N$^+$ ionic abundances.
The electron number density $N_e$ is derived from the [S {\sc ii}] $\lambda$6717/6731
flux ratio when it is available, and is set to 10 cm$^{-3}$ when it is not.
The total element abundances are obtained using ionization correction factors
given by \citet{I06a}.
The element abundances are given in Table \ref{tab4}.

Within the errors, the oxygen abundance derived for DDO 68-3
from the APO spectrum is in agreement with the one derived from 
the MMT spectrum. Note, however, that the
abundance determination from the lower signal-to-noise ratio and lower
resolution APO spectrum is much more uncertain. The derived values 12 + log O/H = 
7.15 $\pm$ 0.04 (DDO 68-3) and 12 + log O/H = 7.16 $\pm$ 0.09 
(DDO 68-4) are in very
good agreement with 12 + log O/H = 7.14 $\pm$ 0.03, 7.13 $\pm$ 0.07 and
7.21 $\pm$ 0.14 derived by \citet{IT07} for regions 1, 2 and 6, respectively.
They are also consistent with the abundance determinations for other 
regions in DDO 68 by \citet{P05} and \citet{P08} from lower signal-to-noise 
ratio spectra. There appears not to be a large abundance variation 
among various H {\sc ii} regions in DDO 68 over a spatial scale of $\sim$ 1 kpc.
The relative abundance homogeneity puts stringent 
constraints on the physics of metal diffusion and mixing processes in this 
object.
The overall conclusion is that we confirm and reenforce the finding by
\citet{IT07} that DDO 68 is the second lowest-metallicity emission-line
galaxy known, after SBS 0335--052W which has 12 + log O/H = 7.12 $\pm$ 0.03
\citep{I05}. This makes the discovery of a LBV star in DDO 68 particularly
important. Being the lowest metallicity LBV known, its study will allow 
to put stringent constraints on evolution models of metal-deficient massive stars.

The oxygen abundance of PHL 293B derived from the VLT spectrum is significantly 
higher than those of the H {\sc ii} regions in DDO 68, being equal to 
12 + log O/H = 7.72 $\pm$ 0.01. But if DDO 68 is excluded, 
it is still considerably lower than 
the oxygen abundances 
of all other galaxies with detected LBVs. 
Within the errors, it is
consistent with 12 + log O/H = 7.66 $\pm$ 0.04 derived by \citet{I07},
from the lower resolution SDSS spectrum. It is also interesting to note that the
nitrogen-to-oxygen abundance ratio log N/O = --1.55 $\pm$ 0.02 is typical
of low-metallicity emission-line galaxies \citep{I06a}. 

\subsection{Broad emission}

\begin{figure*}
\figurenum{5}
\hbox{\includegraphics[angle=-90,width=1.0\linewidth]{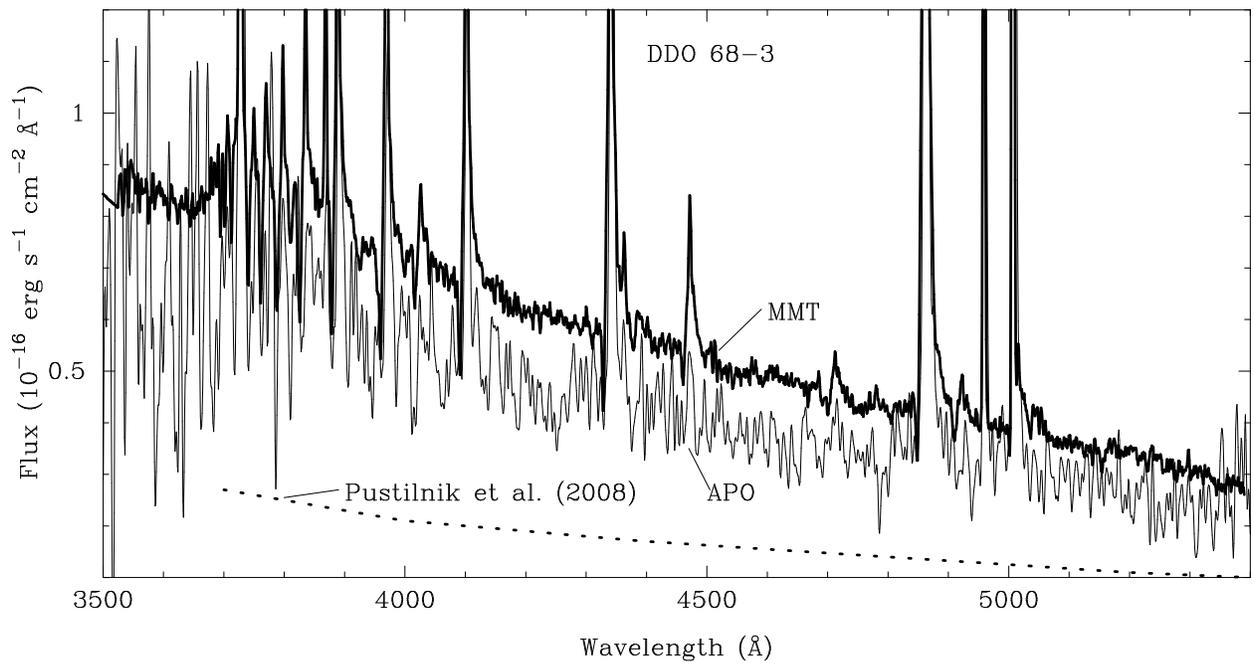} 
}
\figcaption{Comparison of the 
MMT spectrum (thick solid line) and of the APO apectrum (thin solid line) 
of DDO 68-3 shows an increase of the continuum level of a factor of $\sim$ 
3.
The dotted line shows the continuum level before 
the LBV outburst \citep[middle panel of Fig.1 in ][]{P08}.
\label{fig5}}
\end{figure*}

Using Gaussian fitting, we have 
measured the fluxes of the broad components of the Balmer H
lines in the spectra of DDO 68-3 and PHL 293B, and of the He {\sc i} 
lines in the spectrum of DDO 68-3 after 
subtraction of the narrow component. The extinction-corrected fluxes of the broad 
H$\alpha$, H$\beta$ and H$\gamma$ lines together with
the fluxes of the narrow [O {\sc iii}] $\lambda$4959 and 5007 nebular lines, 
derived from the MMT observations of DDO 68-3 and from the VLT observations of 
PHL 293B, are 
shown in Table \ref{tab5}. We also present in the same table the fluxes of the 
same lines, derived from the APO spectrum for DDO 68-3 and from the SDSS spectrum
for PHL 293B. Comparison of the data for DDO 68-3 obtained at different 
epochs (2 February 2008 for the APO spectrum and 28 March 2008 for the MMT spectrum) 
reveals a flux increase of a factor of $\sim$ 2 for H$\beta$ over the course of nearly  
2 months. This is likely a 
real change as the fluxes of the nebular emission lines 
[O {\sc iii}] $\lambda$4959, 5007 remain approximately constant over this period. 
This conclusion is
further supported by the observations of DDO 68-3 by \citet{P08}. 
They derived
a broad H$\beta$ flux of 5.46$\times$10$^{-16}$ erg s$^{-1}$ cm$^{-2}$ on
11 January 2008, or $\sim$ 2 times lower than the broad H$\beta$ flux observed 
on 2 February 2008 at the APO and $\sim$ 4 times lower than the one observed 
on 28 March 2008 at the MMT, 2.5 months later. 
It thus appears that the LBV star 
in DDO 68 is experiencing a significant increase of emission line brightness.
In Fig. \ref{fig5}, we compare the MMT spectrum (thick solid line)
with the APO spectrum (thin solid line). The continuum level before the 
outburst as observed by 
\citet{P08} (their Fig. 1, middle panel), is shown by a dotted line. 
It can be seen that the line flux 
brightening is accompanied by an increase in the continuum level. 
The continuum level of the MMT spectrum at $\lambda$4686, 
the average wavelength of 
the $g$ band, is higher by a factor of $\sim$ 3 than the continuum level of 
the \citet{P08} spectrum, taken 
before the outburst. Adopting an apparent magnitude $g$ = 20.17 for DDO 68-3
before the eruption (Table \ref{tab2}) and a 
distance modulus of 30 mag \citep[which corresponds
to a distance of 10 Mpc; ][]{P08}, the $g$ absolute brightness of 
the LBV star in DDO 68-3 is $\sim$
 $-$11 mag on 28 March 2008. If we adopt the smaller distance of 6.24 Mpc 
given in the  
NED\footnote{NASA/IPAC Extragalactic Database (NED)
is operated by the Jet Propulsion Laboratory, California Institute of 
Technology, under contract with the National Aeronautics and Space 
Administration.}, then
the $g$ absolute magnitude of the LBV star is $\sim$ $-$10 mag. 
In either case, the absolute brightness 
is in the range of those of 
known LBVs \citep{HD94}.
Note that \citet{P08} also found an increase in the continuum level of 
a factor of $\sim$ 2  
between 11 January 2008
and 4 February 2008 (their Fig.2).
We plan to obtain both images and spectra of DDO 68 at regular intervals to monitor 
the LBV's brightness evolution in time.

As for PHL 293B, the fluxes of the broad H lines appear also at first glance 
to vary with  
time: they decrease by a factor of
nearly 2 between the SDSS (22 August 2001) and VLT (8 November 2008) observations. 
However, a more detailed look casts doubt on that interpretation: 
the fluxes of
the nebular [O {\sc iii}] $\lambda$4959, 5007 emission lines also decrease by a 
similar factor
of $\sim$ 1.5. Therefore, these variations may not be real and 
could be caused by aperture differences  
between the VLT and SDSS observations. A narrow slit of 1\arcsec\ width was used in 
the VLT observations while a round 3\arcsec\ aperture was used during the SDSS observations.
In contrast to the VLT slit, the larger SDSS aperture probably included 
all the light from the LBV star.

The H$\alpha$ luminosities of the LBV stars (Table \ref{tab5}) have been calculated 
adopting distances from the NED $D$ = 6.24 Mpc and 22.7 Mpc for DDO 68 and 
PHL 293B, respectively. These distances have been obtained from the radial velocities
corrected for Virgo Infall with a Hubble constant of 73 km s$^{-1}$ Mpc$^{-1}$.
The derived luminosity $L$(H$\alpha$) =
9.4$\times$10$^{36}$ erg s$^{-1}$ (Table \ref{tab5}) for the LBV in DDO 68 from the 
APO spectrum is 
significantly lower than $L$(H$\alpha$) = 3.4$\times$10$^{38}$ erg s$^{-1}$ of 
the LBV in PHL 293B as derived from the SDSS spectrum. 
We have no data for the H$\alpha$ flux for the DDO 68 star from the MMT 
observations as they did not cover the red wavelength range. 
However, if we accept that the H$\alpha$ flux has brightened by a factor of $\sim$ 2 
in the period between the APO and MMT oservations as the 
H$\beta$ flux and that the distance to DDO 68 can be
as high as 10 Mpc \citep[see discussion in ][]{P08}, we conclude that 
the H$\alpha$ luminosity of the LBV in DDO 68 can be as high as 
5$\times$10$^{37}$ erg s$^{-1}$ on 28 March 2008. Then the H$\alpha$
luminosities of the LBVs in DDO 68 and PHL 293B compare well with
$L$(H$\alpha$) $\sim$ 10$^{38}$ erg s$^{-1}$ of the V1 star in NGC 2363,
the third lowest-metallicity LBV known \citep{D01} after those   
in PHL 293B and DDO 68.

\subsection{Stellar winds}

In Table \ref{tab6}, we show the terminal velocities $v_\infty$ of the stellar winds 
associated with the outbursts and
the FWHMs of the broad H and He {\sc i} lines of the DDO 68 and PHL 293B 
LBVs, as derived from the 
high-resolution MMT and VLT spectra. The stellar wind terminal velocity $v_\infty$ is 
derived from the wavelength
difference between the blue absorption minimum and the broad emission line 
maximum. For both LBVs, the $v_\infty$'s and FWHMs derived from different lines
are in very good agreement. Our derived FWHMs for the LBV in DDO 68-3 of 
$\sim$ 1000 km s$^{-1}$ are in 
good agreement with those derived by \citet{P08}. However, while our  
FWHMs are consistent from line to line
(Table \ref{tab6}), those derived by \citet{P08} show a large spread, 
mainly because of their lower signal-to-noise ratio spectra.
The FWHMs measured for the LBV in PHL 293B are $\sim$ 600 km s$^{-1}$,
lower than those of 
the LBV in DDO 68 by a factor of $\sim$ 1.7.
However, the FWHMs of both LBVs are 
significantly broader than those observed in the spectra of high-metallicity 
LBVs which are about 100-200 km s$^{-1}$ \citep[e.g. ][]{L94}.
These differences are probably not all due to metallicity effects. 
Part of the FWHM 
differences may be explained by the fact that the 
high-metallicity LBVs are observed during their quiescent 
phase, while our low-metallicity LBVs are 
 likely observed during their eruptive phase.

The same high velocities are found when $v_\infty$ is considered.    
Indeed, one of the most striking features of low-metallicity LBVs 
is that the terminal velocities of their winds are significantly higher 
than those of higher metallicity LBVs.    
The range of terminal velocities in
higher metallicity LBVs, including V1 in NGC 2363, is 200 -- 400 km s$^{-1}$.
For example, the galactic 
LBV P Cygni has a $v_\infty$ of only 185 km s$^{-1}$ \citep{C04}, 
while in the low-metallicity LBVs $v_\infty$ is $\sim$ 800 km s$^{-1}$. 
Probably,
selection effects may play a role here: generally, emission-line galaxies 
are observed spectroscopically at  
low resolution, so that their spectra do not allow the   
finding of LBVs with low $v_\infty$. In particular, with the 
spectral resolution of $\sim$ 2-3 \AA\ of the SDSS spectra, it is unlikely that we 
can use them to find LBV
with terminal velocities $\la$ 200 km s$^{-1}$.
On the other hand, no high-metallicity LBV was ever found to have 
high terminal velocities. Most likely, metallicity does play a role 
and the stellar winds 
in low-metallicity LBVs are driven by a mechanism which is 
different from the
one operating in high-metallicity LBVs, thought to be the 
radiation pressure due to the permitted lines of heavy elements. 
\citet{SO06} have considered a 
continuum-driven mechanism which could be more efficient at low metallicities
than the line-driven mechanism. It is likely that winds in low-metallicity
LBVs are more transparent and less dense, which results in higher terminal 
velocities. 

New
high spatial resolution spectral and photometric observations are necessary
to shed light on the physical properties and evolution of low-metallicity
LBVs. They will strongly constrain evolutionary
models of low-metallicity massive stars on the post-main-sequence stages of their
evolution. This will in turn increase our understanding of the evolution of 
the first generation of massive stars, when the gas was pristine. 

\section{CONCLUSIONS}

We have studied here the broad line emission of two luminous blue variable (LBV) stars
discovered in two low-metallicity blue compact dwarf (BCD) galaxies, DDO 68 and PHL 293B.
We have arrived at the following conclusions:

1. The oxygen abundances in region 3 of DDO 68 (DDO 68-3) 
and in PHL 293B are respectively 
12+log O/H = 7.15 $\pm$ 0.04 and 12+log O/H = 7.72 $\pm$ 0.01. These two BCDs
are thus the lowest-metallicity galaxies with detected LBV stars. 
PHL 293B is also the most distant galaxy where a LBV star is seen.

2. Photometric observations of DDO 68-3 show that the outburst in
its LBV star has occurred sometime between 9 February 2007 and 11 January 2008.

3. Broad H and He {\sc i} emission lines with P Cygni profiles 
are seen in the spectrum of the LBV star in DDO 68. On the 
other hand, only H broad emission lines are detected in the spectrum
of the LBV in PHL 293B. In both LBVs, no heavy element emission line  
such as Fe {\sc ii} 
was detected, presumably because of their low metallicities. 
The broad H$\alpha$ luminosities in both LBVs
compare well with the one in the higher-metallicity LBV in NGC 2363
\citep{D97,D01,P06}. We find a strong increase of the H$\alpha$
luminosity of the LBV in DDO 68: during the period
from 11 January 2008 to 28 March 2008, it brightened by a factor of $\sim$ 4.

4. One of the most striking features of both low-metallicity LBVs is the 
high terminal velocities $v_\infty$ of their stellar winds. 
Their $v_\infty$ are $\sim$ 800 km s$^{-1}$,   
several times greater than the $v_\infty$ of $\sim$ 100 -- 200 km s$^{-1}$ in 
their high-metallicity counterparts. 
Probably,
metallicity plays a role and winds in low-metallicity LBVs are driven
by a mechanism which differs from the one operating in high-metallicity LBVs. 

\acknowledgements
George Privon, George Trammell and David Whelan kindly obtained the 
spectrum of PHL 293B for us.
Y.I.I. is grateful to the staff of the Astronomy Department at the 
University of Virginia for their warm hospitality. T.X.T. thanks 
the hospitality of the Institut d'Astrophysique de Paris. 
We thank the financial support of National Science Foundation
grant AST02-05785. The 3.5 APO time was available thanks to a grant from the
Frank Levinson Fund of the Silicon Valley Community Foundation
to the Astronomy Department of the University of Virginia.
Funding for the Sloan Digital Sky Survey (SDSS) and SDSS-II has been 
provided by the Alfred P. Sloan Foundation, the Participating Institutions, 
the National Science Foundation, the U.S. Department of Energy, the National 
Aeronautics and Space Administration, the Japanese Monbukagakusho, the 
Max Planck Society, and the Higher Education Funding Council for England. 

\clearpage

\clearpage

\begin{deluxetable}{lcccccc}
  \tablecolumns{6}
  \tablewidth{0pc}
  \tablecaption{Journal of Observations \label{tab1}}
  \tablehead{\colhead{Telescope} &\colhead{Object} & \colhead{Date}
&\colhead{Wavelength range}&\colhead{Slit}& \colhead{Exposure} 
& \colhead{Airmass}
}
  \startdata
2.1m KPNO  & DDO 68  & 09 Feb 2007 & $g$-band    & \nodata                       & 1800  & 1.01    \\
2.1m KPNO  & DDO 68  & 09 Feb 2007 & $i$-band    & \nodata                       & 1800  & 1.05    \\
2.5m Sloan      & DDO 68  & 16 Apr 2004 & $g$-band    & \nodata                       &   54  & \nodata \\
2.5m Sloan      & DDO 68  & 16 Apr 2004 & $i$-band    & \nodata                       &   54  & \nodata \\
3.5m APO   & DDO 68  & 02 Feb 2008 & 3600-9000   &1\farcs5$\times$300\arcsec     & 2700  & 1.70    \\
6.5m MMT   & DDO 68  & 28 Mar 2008 & 3200-5200   &1\farcs5$\times$180\arcsec     & 4500  & 1.08    \\
2.5m Sloan      & PHL 293B& 22 Aug 2001 & 3800-9200   & 3\arcsec (round)              & 2400  & 1.25    \\
8m VLT     & PHL 293B& 08 Nov 2002 & 3100-4500   &1\arcsec$\times$8\arcsec (blue)& 3300  & 1.10    \\
           &         &             & 4800-6800   &1\arcsec$\times$12\arcsec (red)& 3300  & 1.10    \\
  \enddata
  \end{deluxetable}

\clearpage

\begin{deluxetable}{cccccccc}
  \tablecolumns{8}
  \tablewidth{0pc}
  \tablecaption{Photometry of DDO 68 \label{tab2}}
  \tablehead{
\colhead{} & \multicolumn{2}{c}{2.1m KPNO}&&\multicolumn{2}{c}{2.5m Sloan}&& \\ \cline{2-3} \cline{5-6}
\colhead{Region}&\colhead{$g$}&\colhead{$i$}&&\colhead{$g$}&\colhead{$i$}&\colhead{$\Delta$$g$\tablenotemark{a}}&\colhead{$\Delta$$i$\tablenotemark{a}}
}
  \startdata
  &\multicolumn{2}{c}{09 Feb 2007}&&\multicolumn{2}{c}{16 Apr 2004} \\ \cline{2-3} \cline{5-6}
1 &19.44$\pm$0.02&20.27$\pm$0.05&&19.41$\pm$0.04&20.46$\pm$0.18&$+$0.03$\pm$0.04&$-$0.19$\pm$0.19\\
2 &20.44$\pm$0.03&21.38$\pm$0.09&&20.40$\pm$0.07&21.05$\pm$0.31&$+$0.04$\pm$0.08&$+$0.33$\pm$0.32\\
3 &20.17$\pm$0.02&20.63$\pm$0.06&&20.23$\pm$0.05&21.03$\pm$0.29&$-$0.06$\pm$0.05&$-$0.40$\pm$0.30\\
4 &20.19$\pm$0.04&21.04$\pm$0.10&&20.18$\pm$0.07&21.66$\pm$0.54&$+$0.01$\pm$0.08&$-$0.62$\pm$0.55\\
5 &19.13$\pm$0.01&19.41$\pm$0.02&&19.08$\pm$0.02&19.54$\pm$0.08&$+$0.05$\pm$0.02&$-$0.13$\pm$0.08\\
6 &20.40$\pm$0.04&20.84$\pm$0.06&&20.38$\pm$0.07&20.40$\pm$0.15&$+$0.02$\pm$0.08&$+$0.44$\pm$0.16
  \enddata
\tablenotetext{a}{magnitude difference between KPNO and SDSS measurements}
  \end{deluxetable}

\clearpage

  \begin{deluxetable}{lrrrr}
  \tablecolumns{5}
  \tablewidth{0pc}
  \tablecaption{Intensities of narrow emission lines
\label{tab3}}
  \tablehead{
  \colhead{\sc Ion}
  &\colhead{100$\times$$I$($\lambda$)/$I$(H$\beta$)}
  &\colhead{100$\times$$I$($\lambda$)/$I$(H$\beta$)}
  &\colhead{100$\times$$I$($\lambda$)/$I$(H$\beta$)}
  &\colhead{100$\times$$I$($\lambda$)/$I$(H$\beta$)}}
  \startdata
  & \multicolumn{4}{c}{Galaxy} \\ \cline{2-5}
  &
 \multicolumn{1}{c}{DDO 68-3}&\multicolumn{1}{c}{DDO 68-3}&
 \multicolumn{1}{c}{DDO 68-4}&
 \multicolumn{1}{c}{PHL 293B} \\ 
 & \multicolumn{1}{c}{(MMT)}&\multicolumn{1}{c}{(APO)}&
 \multicolumn{1}{c}{(MMT)}&
 \multicolumn{1}{c}{(VLT)} \\  \tableline
3727 [O {\sc ii}]                 &  50.62 $\pm$   1.04 &  57.83 $\pm$   2.59 &  94.10 $\pm$   1.70 &  50.21 $\pm$   0.84 \\
3868 [Ne {\sc iii}]               &  10.07 $\pm$   0.40 &   6.62 $\pm$   0.89 &   5.22 $\pm$   0.09 &  48.49 $\pm$   0.77 \\
3889 He {\sc i} + H8              &  17.74 $\pm$   0.62 &  19.81 $\pm$   1.91 &  18.13 $\pm$   1.20 &  21.27 $\pm$   0.41 \\
3968 [Ne {\sc iii}] + H7          &  16.35 $\pm$   0.59 &  21.39 $\pm$   1.78 &  15.72 $\pm$   1.13 &  31.70 $\pm$   0.54 \\
4101 H$\delta$                    &  26.73 $\pm$   0.68 &  28.39 $\pm$   1.67 &  25.56 $\pm$   1.14 &  27.00 $\pm$   0.46 \\
4340 H$\gamma$                    &  49.34 $\pm$   0.98 &  46.39 $\pm$   1.70 &  47.71 $\pm$   1.27 &  48.25 $\pm$   0.74 \\
4363 [O {\sc iii}]                &   2.82 $\pm$   0.25 &   2.28 $\pm$   0.53 &   1.64 $\pm$   0.38 &  14.55 $\pm$   0.27 \\
4471 He {\sc i}                   &   3.28 $\pm$   0.25 &   \nodata~~~~       &   \nodata~~~~       &   3.75 $\pm$   0.10 \\
4861 H$\beta$                     & 100.00 $\pm$   1.70 & 100.00 $\pm$   2.30 & 100.00 $\pm$   1.95 & 100.00 $\pm$   1.44 \\
4959 [O {\sc iii}]                &  39.56 $\pm$   0.71 &  27.37 $\pm$   0.83 &  23.79 $\pm$   0.72 & 195.13 $\pm$   2.79 \\
5007 [O {\sc iii}]                & 115.29 $\pm$   1.84 &  92.85 $\pm$   2.03 &  68.36 $\pm$   1.40 & 584.35 $\pm$   8.33 \\
5876 He {\sc i}                   &   \nodata~~~~       &   \nodata~~~~       &   \nodata~~~~       &  10.07 $\pm$   0.17 \\
6563 H$\alpha$                    &   \nodata~~~~       & 276.69 $\pm$   5.77 &   \nodata~~~~       & 277.28 $\pm$   4.31 \\
6583 [N {\sc ii}]                 &   \nodata~~~~       &   \nodata~~~~       &   \nodata~~~~       &   1.82 $\pm$   0.06 \\
6678 He {\sc i}                   &   \nodata~~~~       &   \nodata~~~~       &   \nodata~~~~       &   2.76 $\pm$   0.05 \\
6717 [S {\sc ii}]                 &   \nodata~~~~       &   4.31 $\pm$   0.30 &   \nodata~~~~       &   5.04 $\pm$   0.11 \\
6731 [S {\sc ii}]                 &   \nodata~~~~       &   3.00 $\pm$   0.27 &   \nodata~~~~       &   3.68 $\pm$   0.08 \\
 $C$(H$\beta$)                    & 0.000               &  0.005              &  0.000              &  0.080              \\
 EW(H$\beta$) \AA                 & 16                  &    85               &    36               &   37                \\
 $F$(H$\beta$)\tablenotemark{a}   & 21.9                &  29.5               &   9.2               &  71.5               \\
 EW(abs) \AA                      & 0.00                &  1.55               &  0.80               &  0.05               
  \enddata
\tablenotetext{a}{in units 10$^{-16}$ erg s$^{-1}$ cm$^{-2}$.}
  \end{deluxetable}

\clearpage

\begin{deluxetable}{lccc}
  \tablecolumns{4}
  \tablewidth{0pt}
  \tablecaption{Element abundances \label{tab4}}
  \tablehead{
\colhead{Object}&\colhead{12+log O/H}&\colhead{log N/O}&\colhead{log Ne/O}
}
  \startdata
DDO 68-3 (MMT)   & 7.15 $\pm$ 0.04 & \nodata           & --0.81 $\pm$ 0.07 \\
DDO 68-3 (APO)   & 7.08 $\pm$ 0.09 & \nodata           & --0.93 $\pm$ 0.22 \\
DDO 68-4 (MMT)   & 7.16 $\pm$ 0.09 & \nodata           & --1.02 $\pm$ 0.25 \\
PHL 293B (VLT)   & 7.72 $\pm$ 0.01 & --1.55 $\pm$ 0.02 & --0.74 $\pm$ 0.02
  \enddata
  \end{deluxetable}

\clearpage

\begin{deluxetable}{lccccccc}
  \tablecolumns{8}
  \tablewidth{0pc}
  \tablecaption{Fluxes and luminosities of the broad emission lines at 
different epochs \label{tab5}}
  \tablehead{
\colhead{Telescope}&\colhead{Date}&\colhead{$I$(H$\gamma$)\tablenotemark{a}}
&\colhead{$I$(H$\beta$)\tablenotemark{a}}
&\colhead{$I$(H$\alpha$)\tablenotemark{a}}&\colhead{$L$(H$\alpha$)\tablenotemark{b}}
&\colhead{$I$(4959)\tablenotemark{a}}
&\colhead{$I$(5007)\tablenotemark{a}}
}
  \startdata
\multicolumn{8}{c}{DDO 68-3} \\ \tableline

APO      &2008-02-07& \nodata & 9.7$\pm$0.5 & 20.2$\pm$0.4   &9.4$\times$10$^{36}$& 8.4$\pm$0.3 & 26.6$\pm$0.4 \\
MMT      &2008-03-28&10.9$\pm$0.4&19.7$\pm$0.3 & \nodata & \nodata             & 8.8$\pm$0.2 & 25.5$\pm$0.4 \\ \tableline
\multicolumn{8}{c}{PHL 293B} \\ \tableline
SDSS     &2001-08-22& \nodata &26.6$\pm$4.5 & 97.8$\pm$6.3          &6.0$\times$10$^{38}$&196.5$\pm$6.8 & 603.8$\pm$20.1 \\
VLT      &2002-11-08& 7.5$\pm$0.2    &16.7$\pm$0.3 & 55.2$\pm$0.6   &3.4$\times$10$^{38}$&139.9$\pm$1.5 & 412.8$\pm$4.0 \\
  \enddata
\tablenotetext{a}{In units 10$^{-16}$ erg s$^{-1}$ cm$^{-2}$. Flux errors 
are derived taking into account photon statistics in non-flux calibrated 
spectra.}
\tablenotetext{b}{In erg s$^{-1}$.}
  \end{deluxetable}

\clearpage

\begin{deluxetable}{lccccc}
  \tablecolumns{6}
  \tablewidth{0pc}
  \tablecaption{Terminal velocities and FWHMs of the broad emission
lines in DDO 68-3 and PHL 293B \label{tab6}}
  \tablehead{\colhead{} & \multicolumn{2}{c}{DDO 68-3 (MMT)}&\colhead{}
& \multicolumn{2}{c}{PHL 293B (VLT)} \\ \cline{2-3} \cline{5-6}
  \colhead{Line}
&\colhead{$v_\infty$\tablenotemark{a}}
&\colhead{FWHM$_{br}$\tablenotemark{a}}&\colhead{}
&\colhead{$v_\infty$\tablenotemark{a}}
&\colhead{FWHM$_{br}$\tablenotemark{a}}
}
  \startdata
6563 H$\alpha$  & \nodata  & \nodata         && 859     & 719$\pm$10     \\
5016 He {\sc i} &  826     &  974$\pm$62     && \nodata & \nodata \\
4861 H$\beta$   &  801     & 1008$\pm$10     && 819     & 649$\pm$12     \\
4713 He {\sc i} &  807     & 1004$\pm$89     && \nodata & \nodata \\
4471 He {\sc i} &  803     & 1053$\pm$54     && \nodata & \nodata \\
4340 H$\gamma$  &  835     & 1122$\pm$20     && 871     & 681$\pm$14     \\
4101 H$\delta$  &  788     & 1137$\pm$33     && 735     & 474$\pm$30    \\
3968 H7         &  762     & 1100$\pm$40     && \nodata & \nodata \\
3888 H8         &  732     & 1015$\pm$31     && \nodata & \nodata \\
  \enddata
\tablenotetext{a}{In km s$^{-1}$.}
  \end{deluxetable}

\clearpage

\clearpage

\clearpage

\clearpage

\clearpage

\end{document}